\documentclass{cimento}

\usepackage{graphicx} 

\title{Status of charged lepton flavour violation search with MEG II\\ experiment}
\author{M.~Meucci \thanks{Email: \texttt{manuel.meucci@roma1.infn.it}} on behalf of the MEG collaboration}
\instlist{\inst{} Dipartimento di Fisica, Universit\`a di Roma La Sapienza - Roma, Italy
\inst{} INFN, Sezione di Roma, Italy}

\begin{document}

\maketitle

\begin{abstract}
The MEG II experiment searches for the Charged Lepton Flavour Violating (CLFV) decay  $\mu\rightarrow e \gamma$. This decay is foreseen by the Standard Model (SM) of Particle Physics at non observable rates through neutrino oscillation. Observing it would be a clear signal of new physics (\textit{e.g.} SUSY-GUT). After publishing the current upper limit to the Branching Ratio (BR) of this decay in 2016, \\BR($\mu^+\rightarrow e^+ \gamma)<4.2\times10^{-13}$, the MEG experiment started the upgrade of its detector in order to increase the sensitivity by a factor 10, starting the MEG II phase. The aim of this work is to report a description of the new detectors design and performances, to present the results obtained in the 2018 pre-engineering run with all the new MEG II detectors installed, and to report on the current status of the experiment and its future prospects.

\end{abstract}

\section{Introduction}
The SM is the most successful theory that describes elementary particles and their interactions. It  has several experimental confirmations, but it is not a complete theory, since it cannot describe all of the phenomena concerning Particle Physics, and it is believed to be a low energy approximation of a more general theory. Many Beyond SM (BSM)  theories have thus been considered, such as the Grand Unification Theory (GUT) and the Supersymmetric Model (SUSY). The lepton flavour is conserved in the SM, and even including neutrino oscillation the CLFV rate is $\sim10^{-54}$, non observable with experiments. Some BSM theories, such as SUSY-GUT, foresee CLFV at an observable rate, with a BR of order $10^{-14}$.\\
The MEG experiment looked for the CLFV $\mu^+\rightarrow e^+ \gamma$ decay using the highest intensity muon beam available in the world, at the Paul Scherrer Institut (Villigen, CH), that can reach an intensity of $\sim10^8\,\mu^+/s$. In 2016 the MEG collaboration published the best upper limit on that decay, BR($\mu^+\rightarrow e^+ \gamma)<4.2\times10^{-13}$~\cite{ref:meg2016}, using data collected in the period 2009-2013. Due to background limitations, the $\mu^+$ beam intensity used in this data taking was not the maximum available from the PSI facility, but was limited to $3\times10^7\,\mu^+/s$.  In 2013 the MEG collaboration proposed an upgrade of the detectors design and performances, able to cope with a $\mu^+$ beam intensity up to $7\times10^7\,\mu^+/s$, with the aim of improving the overall experiment sensitivity by one order of magnitude~\cite{ref:megiiup}.

\section{New detectors for MEG II}
\begin{figure}[h!]
\centering
\includegraphics[scale=.33]{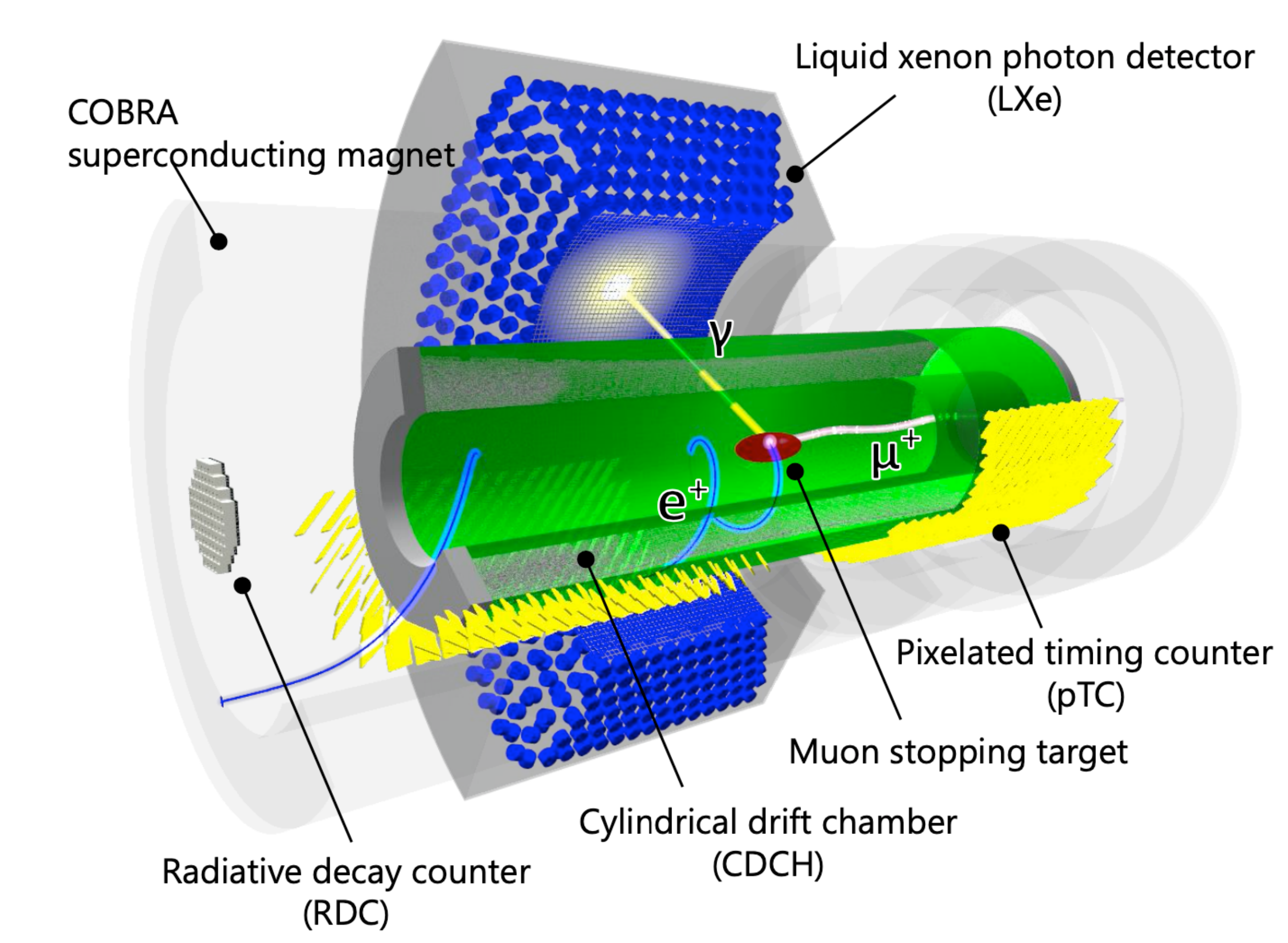}
\caption{Experimental layout of the MEG II experiment.}
\label{fig:meg2}
\end{figure}
The new detectors of the MEG II experiment are reported in fig.~\ref{fig:meg2}~\cite{ref:megii}. The positron detector is a magnetic tracker, composed by a magnet (COBRA, COnstant Bending RAdius) that generates a non homogeneous magnetic field, an ultra light and high precision tracker (CDCH, Cylindrical Drift CHamber), and a fast time detector (pTC, pixelated Timing Counter). The photon detector is an electromagnetic calorimeter (LXe, Liquid Xenon) read by both Photo Multiplier Tubes (PMT) and Multi-Pixel Photon Counters (MPPC). A Radiative Decay Counter (RDC) is also used to tag one type of background events, thus improving the background rejection.

\subsection{Signal and background}
The experimental constraints of MEG II are driven by the signal and background characteristics of the $\mu^+\rightarrow e^+ \gamma$ decay. The muon is slowed down inside a target and decays at rest, so the positron and the photon are emitted back to back with a energy of $E_{e^+}=E_{\gamma}=m_{\mu}/2=52.8$ MeV. There are two main sources of background for this experiment:
\begin{enumerate}
\item the Radiative Muon Decay (RMD) $\mu^+\rightarrow e^+\nu_e\bar{\nu}_\mu\gamma$ when the two neutrinos have low energy;
\item the accidental coincidence of a photon coming from electron-positron annihilation in the detector material (AIF, Annihilation In Flight) or from RMD and a muon coming from the Michel decay $\mu^+\rightarrow e^+\nu_e\bar{\nu}_\mu$.
\end{enumerate}
The number of signal events is proportional to the muon beam intensity and the number of background events is proportional to its square power. It is thus possible to increase the beam intensity only if the detectors resolutions are good enough to keep the background manageable. This reason, and the fact that the AIF must be kept as low as possible (this is achieved reducing the material in the detector, which helps also to reduce multiple scattering), lead to a redesign of the MEG detectors. Thanks to the improvements of MEG II it will be possible to use a $7\times10^7\,\mu^+/s$ muon beam intensity, 2.5 times higher than in MEG.

\subsection{Photon detector}
\begin{figure}[h!]
\centering
\includegraphics[scale=.26]{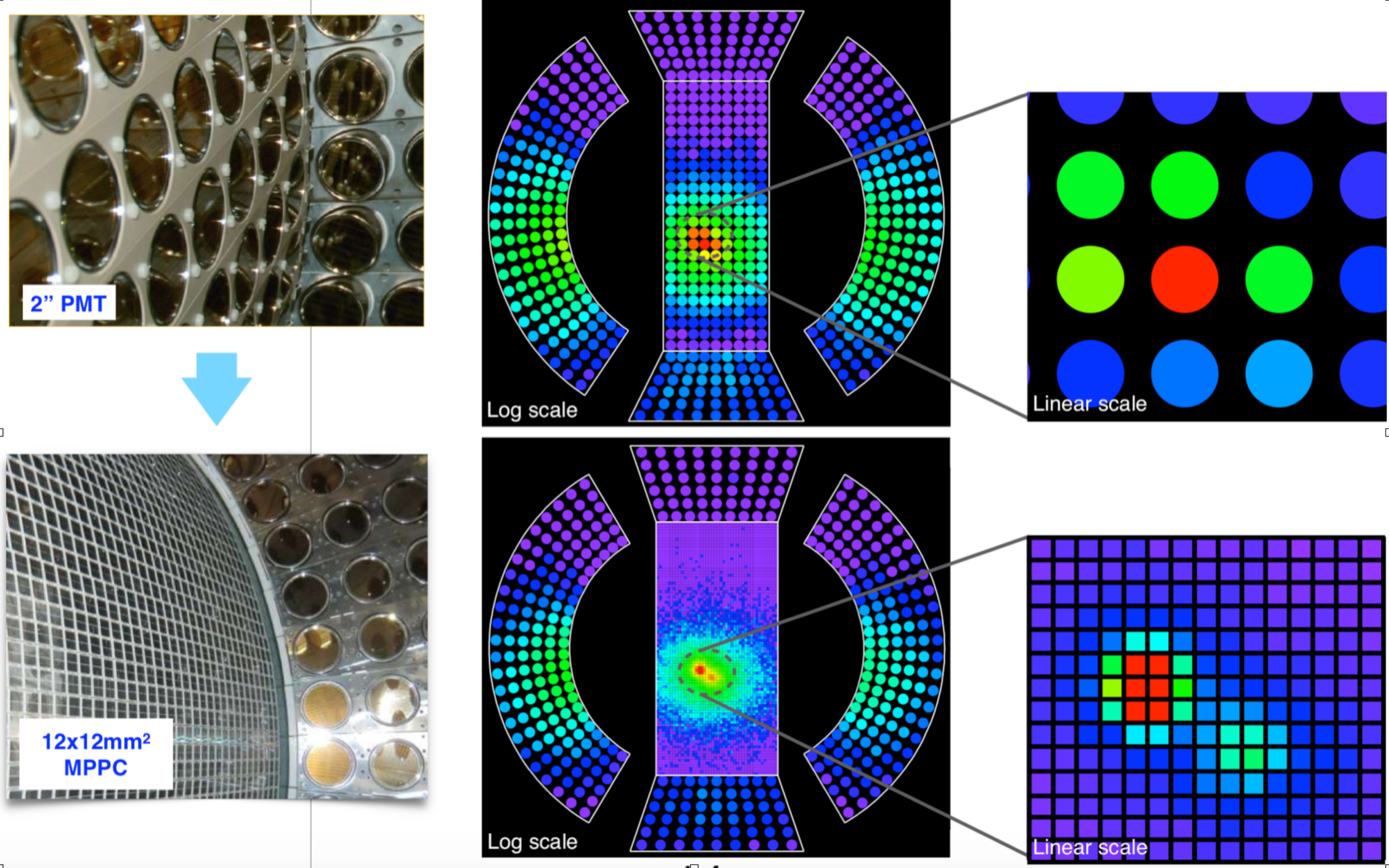}
\caption{ Top: MEG LXe calorimeter internal face covered by PMTs. Bottom: MEG II  LXe calorimeter internal face covered by MPPCs. The image shows both a picture and an example of event display of the detector.}
\label{fig:lxe}
\end{figure}

\begin{figure}[h!]
\centering
\includegraphics[scale=.3]{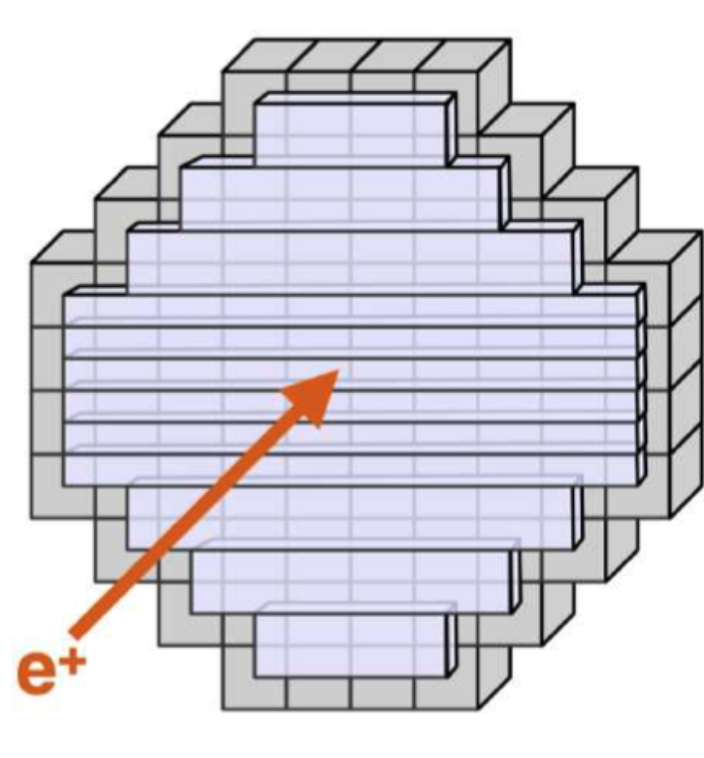}
\caption{Schematic view of the RDC detector.}
\label{fig:rdc}
\end{figure}

The MEG II photon detector is a homogeneous electromagnetic calorimeter that uses Liquid Xenon (LXe) as active material. The scintillation photons emitted by the LXe are detected by 668 PMTs and 4092 MPPCs, sensitive to the VUV light. The concept of this detector is the same of the MEG calorimeter, except for its internal face, previously covered by PMTs and now covered by the 4092 MPPCs, that are insensitive to the magnetic field and have less material budget. Figure~\ref{fig:lxe} shows a picture of the internal face of the detector, along with an example of event display.
\\
The calorimeter can measure the photon characteristics with a precision of $\sigma(E_\gamma)=1.0\%$ (up to 1.1\% depending on the conversion depth) on the energy, $\sigma(t_{e\gamma})=84\,$ps on the timing and $\sigma(u_\gamma/v_\gamma/w_\gamma)=2.6/2.2/5\,$mm on the position. The overall detector efficiency is $\varepsilon(\gamma)=69\%$.\\
An additional detector has been developed for MEG II, with the objective of reducing the RMD background, thus improving the expected experiment sensitivity at 90\% CL from $6\times10^{-14}$ to $4.3\times10^{-14}$. Figure~\ref{fig:rdc} shows a schematic view of this detector, the RDC. It uses plastic scintillators coupled to SiPMs for positron timing, and LYSO crystals coupled to SiPMs for positron energy. The detector is placed along the beamline and after the target, so that it can detect low energy positron coming from RMD and coincident in time with photons, that are a source of background for the LXe.

\subsection{Positron tracker}
The positron tracker for MEG II is completely different from the old one. It has a new low mass single volume 
CDCH for momentum measurement and tracking, and a new pixelated Timing Counter (pTC) for timing. The COBRA magnet is the same one used in MEG.
\begin{figure}[h!]
\centering
\includegraphics[scale=.28]{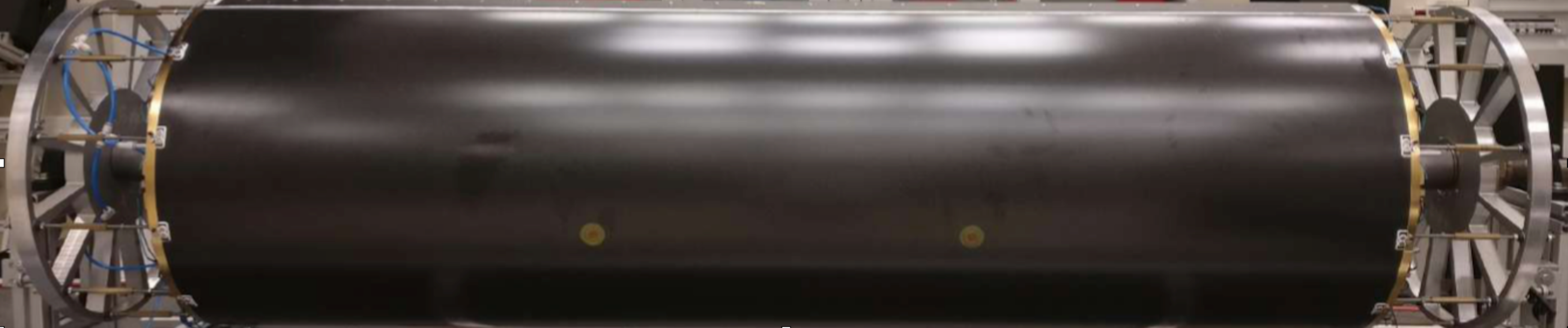}
\caption{Picture of the CDCH closed by a thin Carbon Fiber foil.}
\label{fig:cdch}
\end{figure}\\
The CDCH (fig.~\ref{fig:cdch}) is filled with a 90:10 He:iC$_4$H$_{10}$ gas mixture. It is composed of 9 layers, each containing 192 50 $\mu$m Al sense wires with Ag coating. The wires follow a stereo geometry that allows to measure the positron position along the beam axis.
\begin{figure}[h!]
\centering
\includegraphics[scale=.27]{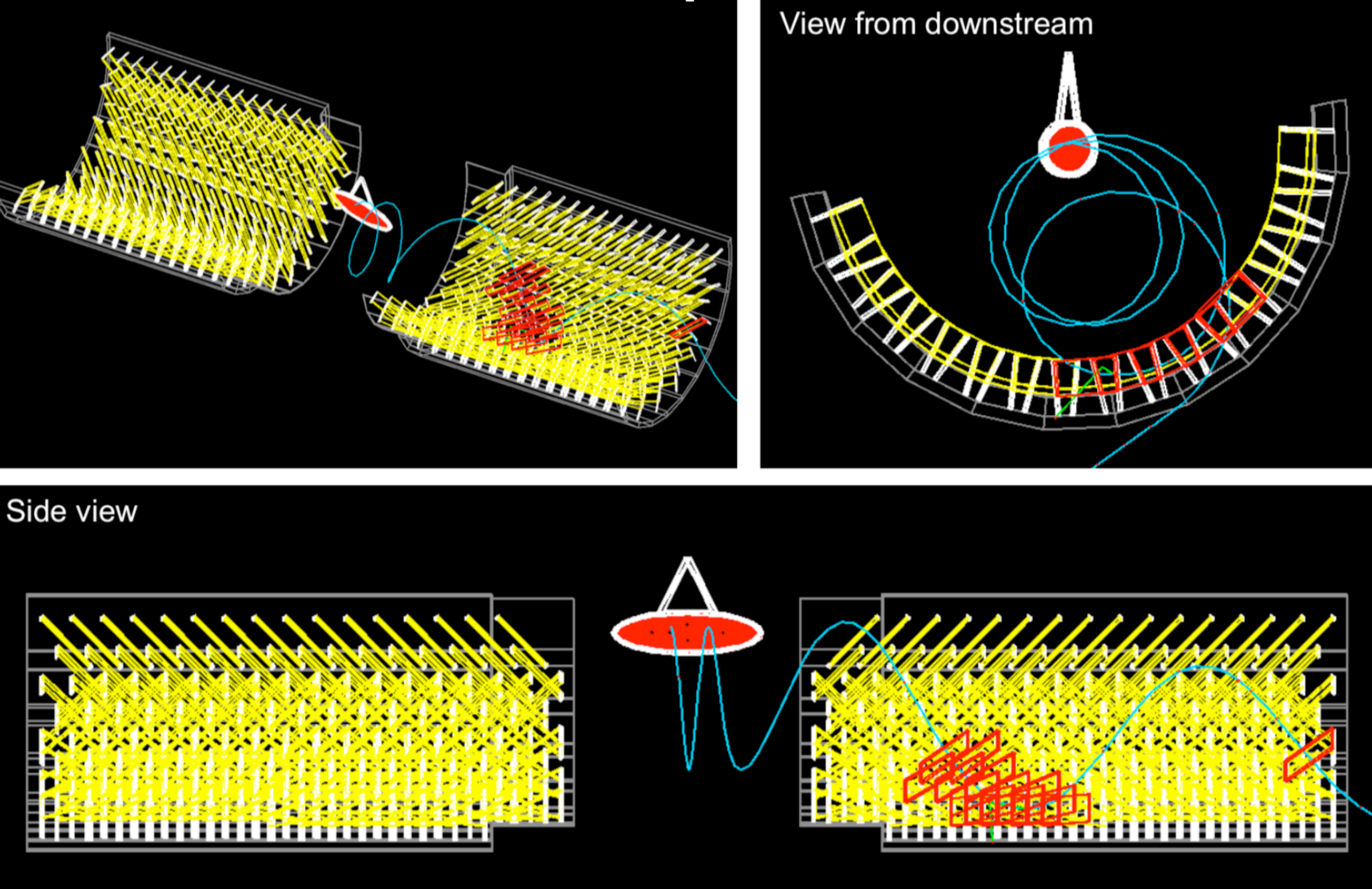}
\caption{Schematic views of the pTC modules.}
\label{fig:ptc}
\end{figure}\\
Figure~\ref{fig:ptc} shows the design of the new pTC. It is composed of 2 main modules, 512 tiles each. Each tile has a plastic scintillator read by SiPMs. The granularity of the detector is increased with respect to the MEG TC, and it is insensitive to the magnetic field, since the old PMTs are replaced by SiPMs.
The CDCH can measure the positron characteristics with a precision of  $\sigma(E_e^+)=130\,$keV on the energy,  $\sigma(\theta_e^+/\phi_e^+)=5.3/3.7\,$mrad on the angle,  $\sigma(z_e^+/y_e^+)=1.6/0.7\,$mm on the position and an efficiency of  $\varepsilon(e^+)=70\%$. \\
The pTC, combined with the LXe time information, allows to measure the photon-positron time with a precision  of $\sigma(t_e^+\gamma)=84\,$ps.

\subsection{TDAQ}
\begin{figure}[h!]
\centering
\includegraphics[scale=.12]{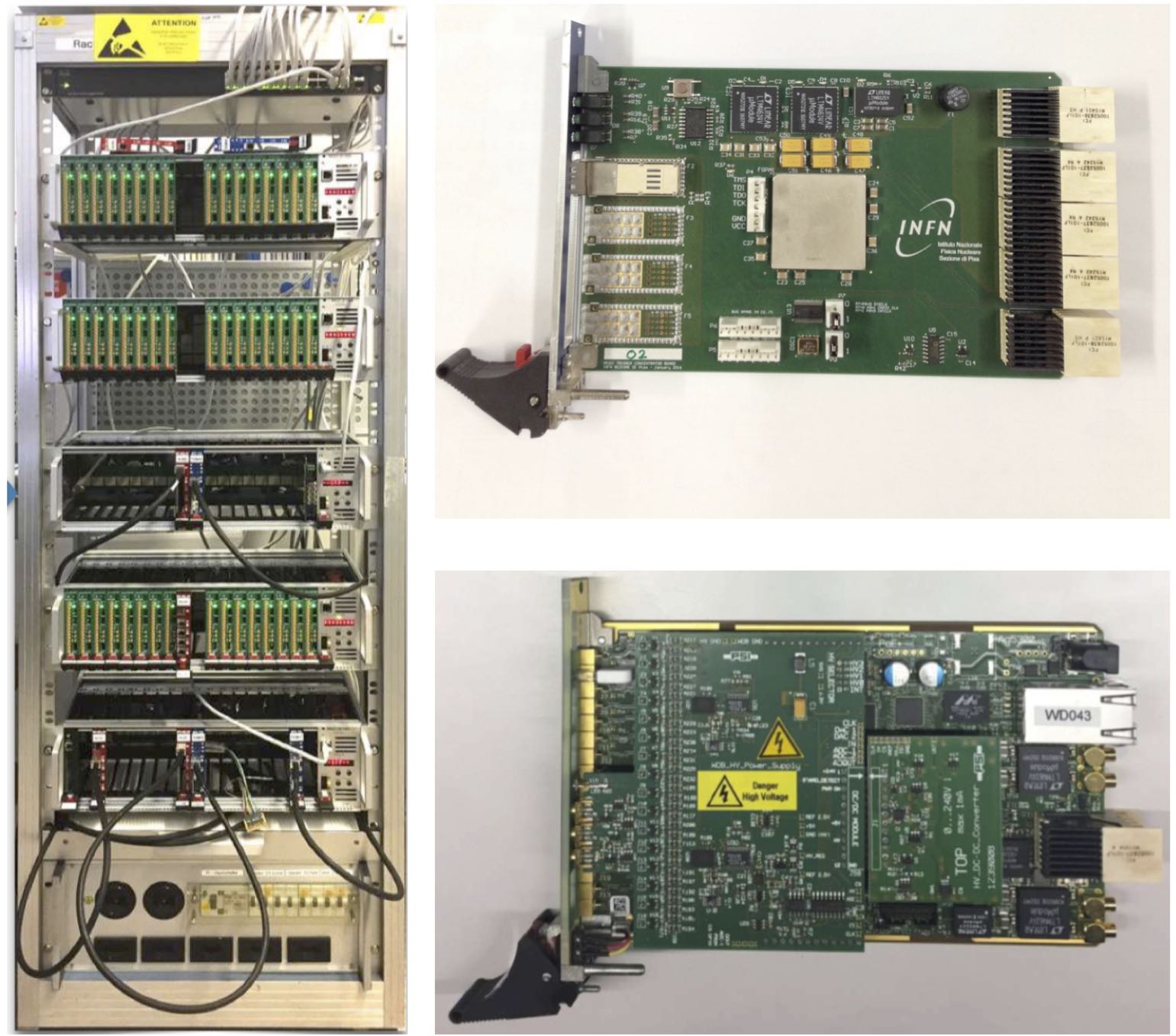}
\caption{Left: MEG II rack. Each rack contains up to 10 crates, each crate contains 16 WDBs and a TCB. Top right: picture of a TCB. Bottom right: picture of a WDB.}
\label{fig:tdaq}
\end{figure}

Since the number of the readout channels in MEG II has increased (MPPCs, more pTC tiles, more CDCH wires) the TDAQ system needed an upgrade. Figure~\ref{fig:tdaq} shows a picture of a TDAQ rack and of the two main boards used in MEG II. The first board is the Trigger Concentrator Board (TCB), that uses a Xilinx Kintex 7 FPGA to build the trigger.
The second board is the WaveDream Board (WDB), that is responsible both for the DAQ, using a custom chip built at PSI, the Domino Ring Sampler (DRS), and for the first steps of the trigger, using a Xilinx Spartan 6 FPGA.

\section{2018 pre-engineering run}
\begin{figure}[h!]
\begin{minipage}{.45\textwidth}
\centering
\includegraphics[scale=.2]{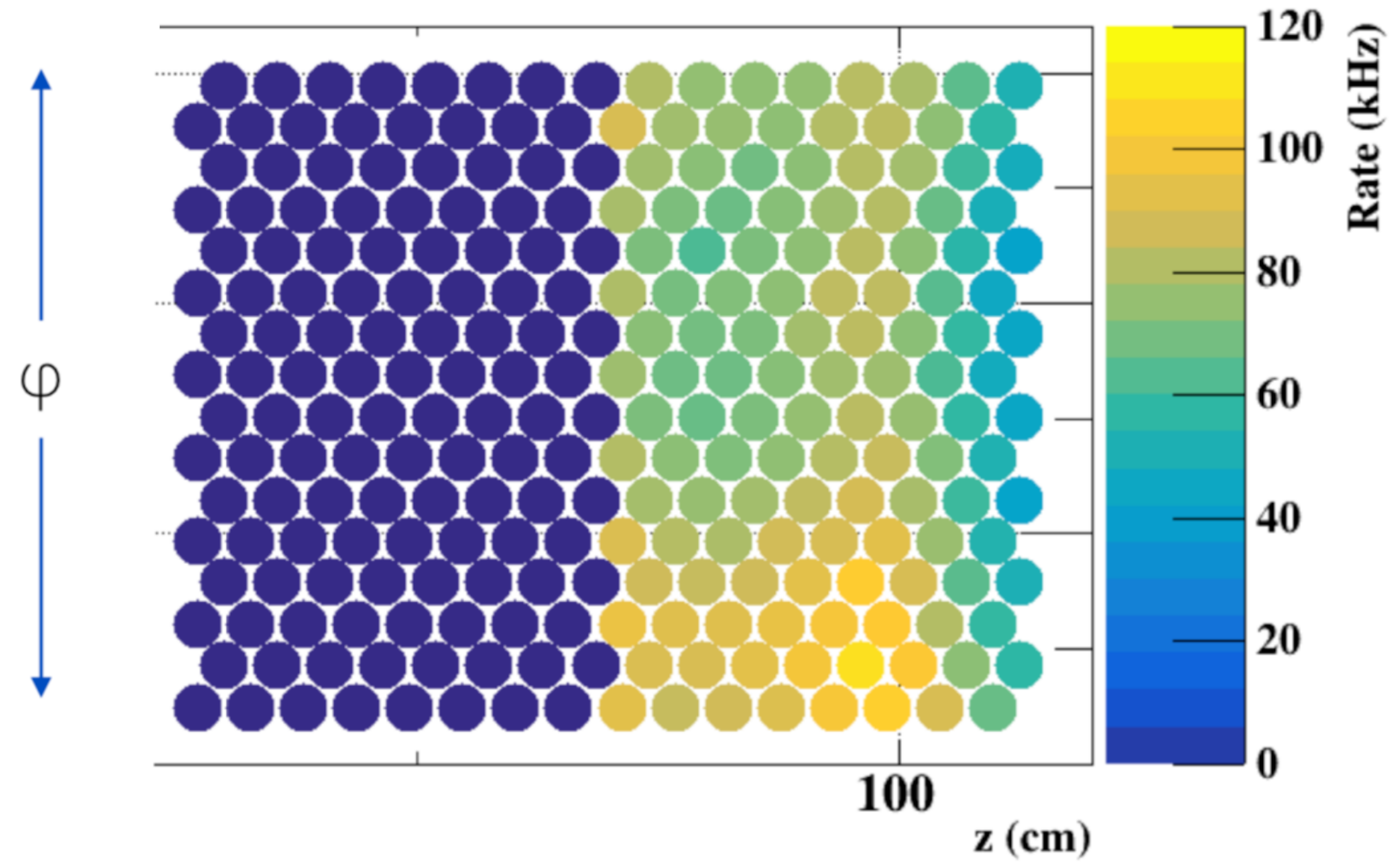}
\end{minipage}
\hspace{.001\textwidth}
\begin{minipage}{.45\textwidth}
\centering
\includegraphics[scale=.225]{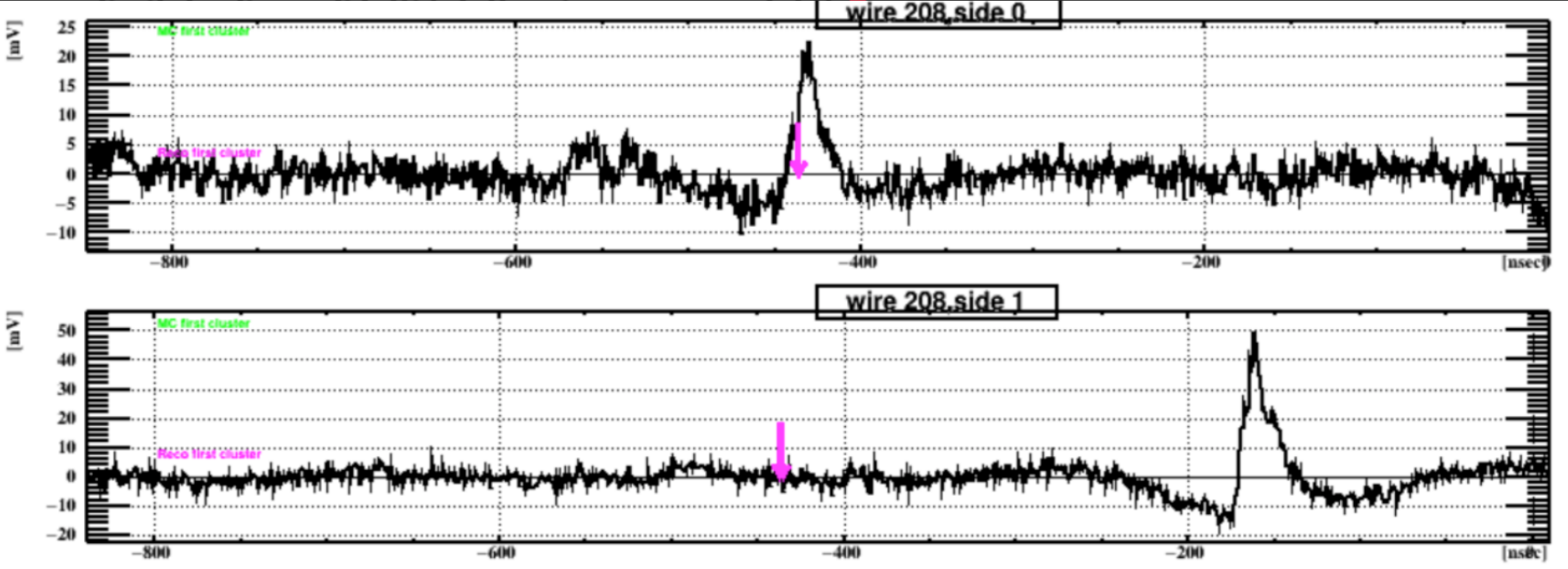}
\end{minipage}
\caption{Left: pTC readout channels with hit rate. Readout only on half of a module. Right: First cosmic ray waveform observed with the CDCH.}
\label{fig:pertccdch}
\end{figure}

In November-December 2018 a MEG II pre engineering run took place at PSI. This was the first data taking with all the new MEG II detectors installed. The number of readout channels was limited, so only a part of each detector has been tested.
\begin{figure}[h!]
\begin{minipage}{.45\textwidth}
\centering
\includegraphics[scale=.22]{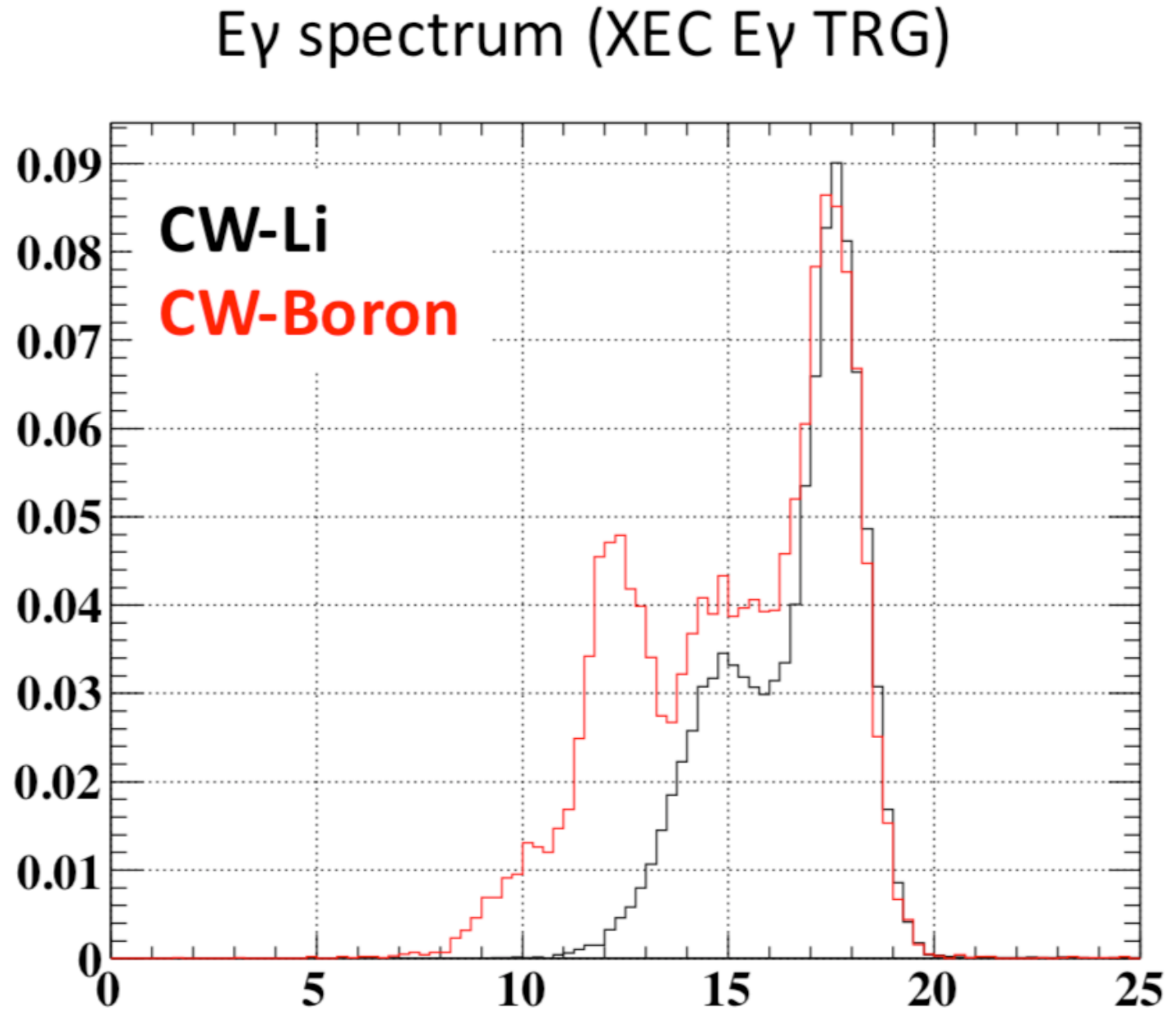}
\end{minipage}
\hspace{.05\textwidth}
\begin{minipage}{.45\textwidth}
\centering
\vspace{11pt}
\includegraphics[scale=.22]{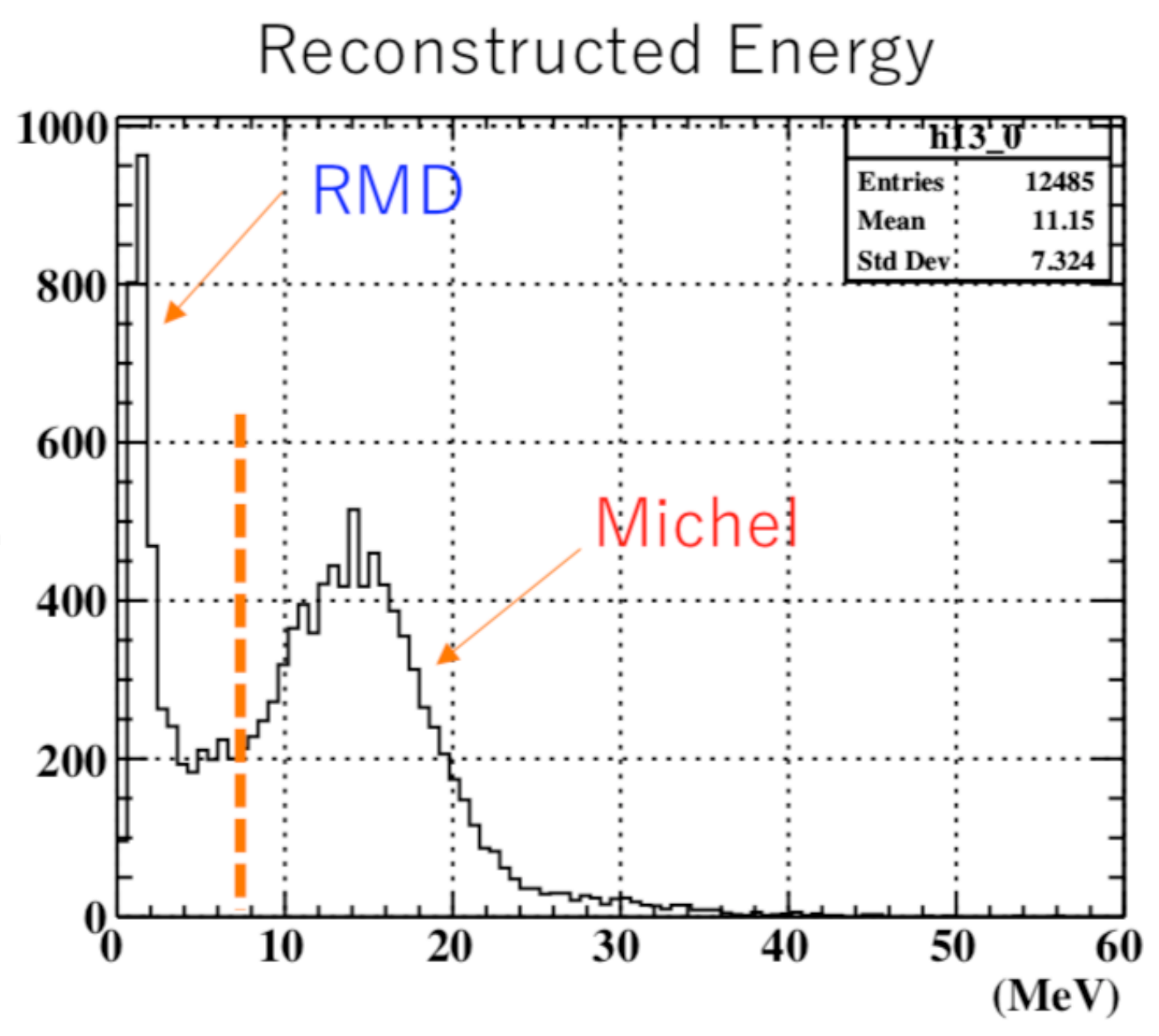}
\end{minipage}
\caption{Left: photon energy spectrum used to calibrate LXe. Right: positron reconstructed energy from RDC.}
\label{fig:perlxerdc}
\end{figure}\\
The RDC main task was the commissioning of the detector. Positrons data have been taken using the LXe calorimeter as a trigger. The reconstructed energy of the positrons collected, reported in fig.~\ref{fig:perlxerdc}, shows that this detector can separate RMD and Michel positrons.
The LXe calorimeter took calibration data using photons coming from a dedicated Li$_2$B$_4$O$_7$ target, on which 1 MeV energy protons, coming from a dedicated Cockroft-Walton (CW) accelerator, impinge. Figure~\ref{fig:perlxerdc} shows the Lithium and Boron calibration lines. The readout was on 10\% of the detector (central MPPCs on the inner face).\\
The pTC took calibration data using both Michel positrons and a dedicated laser light. Only 1/4 of the detector was read out (256 tiles), as shown in fig.~\ref{fig:pertccdch}.
The CDCH took cosmic rays data at different HV and muon beam data at different beam intensities. The purpose was to find a suitable working point for the detector, since it was its first operation in the experiment. One of the first waveform acquired is shown in fig.~\ref{fig:pertccdch}. The mechanical structure of the CDCH was satisfactory, but some electrostatic instability showed up during the data taking. 
The readout was on 192 channels, shared by the three outermost layers of the detector.

\section{Conclusion}
The 2018 pre-engineering run showed how the experiment can operate succesfully with all the detectors installed. The CDCH electrostatic instability has been inspected and succesfully resolved after the run. The MEG II physics data taking will start soon, as all the TDAQ electronics is almost ready.


\begin{thebibliography}{0}
\bibitem{ref:meg2016}
\BY{Baldini A. M. \textit{et al.}}
\IN{Eur. Phys. J. C}{76}{8},\SAME{434}{2016}. doi:10.1140 epjc/s10052-016-4271-x, arXiv:1605.05081
\bibitem{ref:megiiup}
\BY{Baldini A. M. \textit{et al.}} \TITLE{MEG upgrade proposal} (R-99-05.2) (2013). arXiv:1301.7225
\bibitem{ref:megii}
\BY{Baldini A. M. \textit{et al.}}
\IN{Eur. Phys. J. C}{78}{5},\SAME{380}{2018}
\end{thebibliography}
\end{document}